\documentclass[review]{elsarticle}
\usepackage{amsmath}
\usepackage{hyperref}
\hypersetup{colorlinks = true, allcolors = blue}
\usepackage[nameinlink,noabbrev]{cleveref}
\usepackage[margin=1in]{geometry}
\usepackage[font=footnotesize,labelfont=bf]{caption}
\usepackage{hhline}
\usepackage{verbatim}
\usepackage{xcolor}

\begin{document}
\sloppy
\begin{frontmatter}

\title{\textbf{Reconfigurable Enhancement of Actuation Forces by Engineered Losses in non-Hermitian Metamaterials}}

\author[1]{Abhishek Gupta}
\author[2]{Arkady Kurnosov}

\author[2]{Tsampikos Kottos\corref{mycorrespondingauthor1}}
\cortext[mycorrespondingauthor1]{Corresponding author}
\ead{tkottos@wesleyan.edu}

\author[1]{Ramathasan Thevamaran\corref{mycorrespondingauthor2}}
\cortext[mycorrespondingauthor2]{Corresponding author}
\ead{thevamaran@wisc.edu}

\address[1]{Department of Mechanical Engineering, University of Wisconsin-Madison, Madison, WI, 53706, USA}
\address[2]{Wave Transport in Complex Systems Lab, Physics Department, Wesleyan University; Middletown, CT-06459, USA}

\begin{abstract}
While boosting signals with amplification mechanisms is a well-established approach, attenuation mechanisms are typically considered an anathema because they degrade the efficiency of the structures employed to perform useful operations on these signals. An emerging alternate viewpoint promotes losses as a novel design element by utilizing the notion of exceptional point degeneracies (EPDs)—points in parameter space where the eigenvalues of the underlying system and the associated eigenvectors simultaneously coalesce. Here, we demonstrate a direct consequence of such eigenbasis collapse in elastodynamics—an unusual enhancement of actuation force by a judiciously designed non-Hermitian metamaterial supporting an EPD that is coupled to an actuation source. Intriguingly, the EPD enables this enhancement while maintaining a constant signal quality. Our work constitutes a proof-of-principle design which can promote a new class of reconfigurable nano-indenters and robotic-actuators. Importantly, it reveals the ramifications of non-Hermiticity in boosting the Purcell emissivity enhancement factor beyond its expected value, which can guide the design of metamaterials with enhanced emission that does not deteriorate signal quality for mechanical, acoustic, optical, and photonic applications.

\end{abstract}


\end{frontmatter}


Unusual wave phenomena associated with phononic crystals and locally resonant metamaterials have been utilized for various engineering applications from tunable acoustic bandgaps \cite{liu2000locally,hussein2014dynamics} to wave rectification \cite{boechler2011bifurcation}, cloaking \cite{zhang2011broadband}, and non-reciprocal and topological wave guides \cite{cha2018experimental,nassar2017non}. On the other hand, the unveiled wealth of underlying mathematical structures of the non-Hermitian wave systems \cite{bender2007making,kato2013perturbation} and their utilization in new technologies have flourished over the last few years. The non-Hermitian notions have influenced various areas of physics and engineering such as optics and photonics \cite{el2018non,miri2019exceptional}, RF and microwaves \cite{thomas2016giant,doppler2016dynamically,assawaworrarit2017robust}, optomechanics \cite{xu2016topological}, acoustics \cite{shi2016accessing,thevamaran2019asymmetric}, physics of cold atoms \cite{peng2016anti}, magnonics \cite{lee2015macroscopic,zhang2017observation}, and most recently elastodynamics \cite{dominguez2020environmentally,fang2021universal,lustig2019anomalous,rosa2021exceptional,li2022experimental,fang2022emergence}. Consequently, new concepts have been developed and realized in these frameworks. Examples include a loss-induced transparency \cite{guo2009observation}, unidirectional invisibility \cite{lin2011unidirectional}, asymmetric transport with frequency purity \cite{thevamaran2019asymmetric}, parity-time-symmetric lasers \cite{hodaei2014parity,feng2014single}, hypersensitive sensors \cite{hodaei2017enhanced,chen2017exceptional,lai2019observation,hokmabadi2019non,kononchuk2022enhanced}, etc. Many of these phenomena are reliant on the existence of EPDs. Their implementation requires judicious design of attenuation (and/or amplification) and impedance profiles that lead to enhanced wave-matter interactions. 

As opposed to many of these previous studies that emphasized the topology of the eigenfrequency surfaces in the parameter space near an EPD, here we exploit phenomena intimately related to the eigenvector coalescence at EPDs \cite{kato2013perturbation} and its ramifications in the shape of the Local Density of States (LDoS). The latter has recently attracted a lot of attention in the optics framework \cite{lin2016enhanced,pick2017general,pick2017enhanced,ren2021quasinormal,franke2021fermi,zhong2021control,khanbekyan2020decay,hashemi2022linear}. Specifically, it was theoretically argued that a direct consequence of such eigenbasis collapse is an anomalous emissivity enhancement of a source when it is brought at the proximity of an environment (e.g., a cavity) featuring EPDs. The underlying physics of emissivity enhancement of a source due to a surrounding environment is known as Purcell effect and has been initially developed in the framework of quantum electrodynamics \cite{purcell1946resonance}. It described the modification of the spontaneous emission rate of a quantum source by changing its environment via manipulation of the local density of states (LDoS) of the quantum system. The latter information is imprinted in the Green’s function that describes the surrounding environment. Interestingly, this same physics is intimately connected with one of the most fundamental rules of quantum mechanics: The Fermi’s Golden Rule (FGR) \cite{dirac1927quantum} which links emission rates to the LDoS of the system. Elucidating the Purcell physics from this angle allows for expanding its applications to various wave systems where cavity manipulation can be employed to tailor the LDoS of the environment to which a source is coupled. Purcell effect has benefitted numerous technologies such as nano-optical spectroscopy, nanolasing, quantum information processing, energy harvesting, emission enhancement of sound and elastodynamic sources, etc \cite{betzig1993single,muhlschlegel2005resonant,khajavikhan2012thresholdless,choy2011enhanced,schmidt2018elastic,fink2013subwavelength,broitman2017indentation}. An investigation, therefore, of non-Hermiticity and EPDs and their ramifications in FGR and Purcell physics constitutes a dual benefit: on a fundamental level, it will provide us a deeper understanding of the connection between engineered impedance and one of the basic rules, i.e., FGR, while on the technological level, it will allow us to exploit the effects of non-Hermitian spectral singularities like EPDs to enhance the emissivity of mechanical sources. 

\begin{figure}[t]
	\centering
	\includegraphics[width=0.9\textwidth]{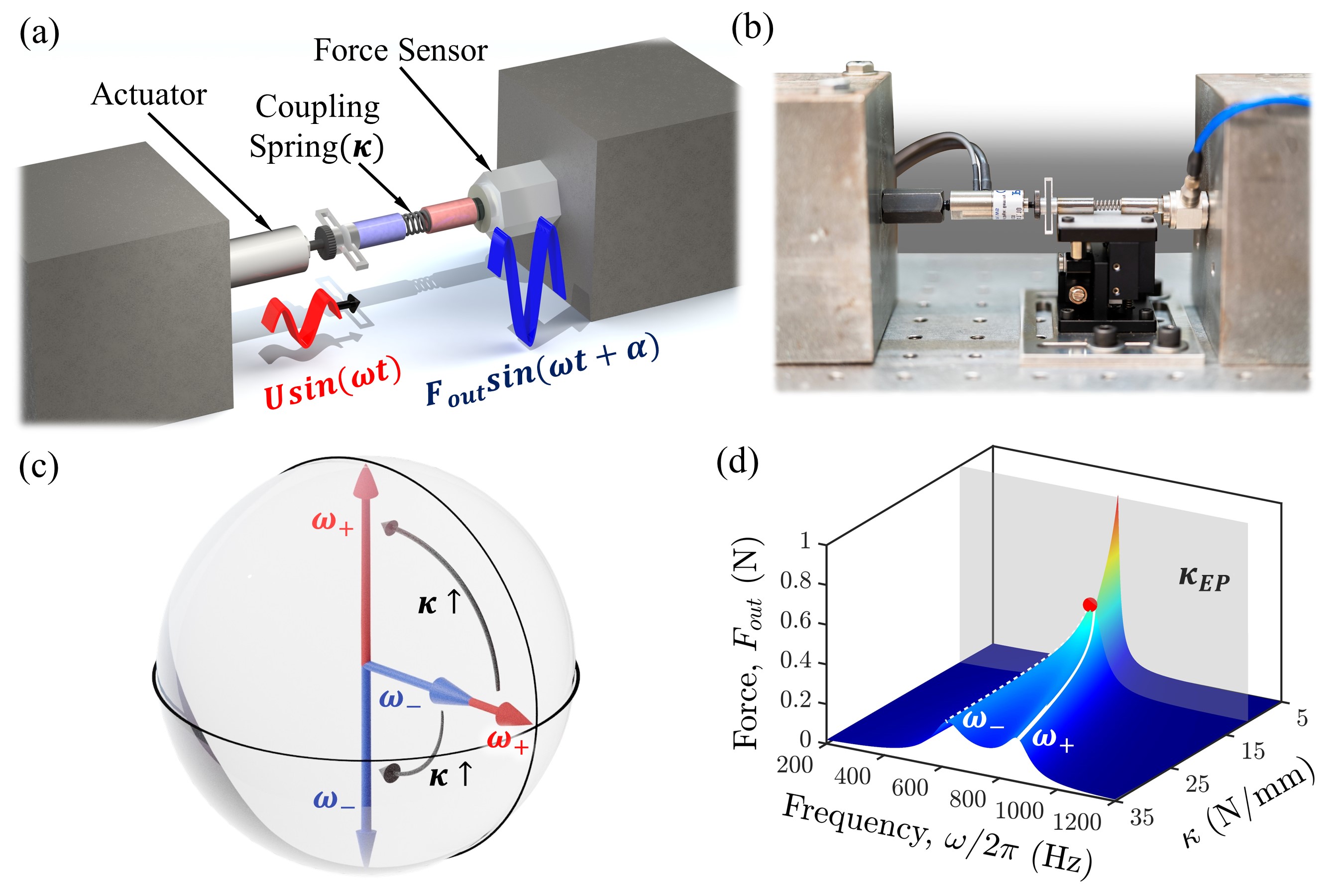}	\caption{ (a) A schematic and (b) a photograph of the experimental setup used in this study showing a non-Hermitian metamaterial consisting of two coupled resonators with one made of a Hookean aluminum spring and the other made of a viscoelastic material (PDMS) with controlled crosslinking. This metamaterial is mounted between a piezoelectric actuator (forced boundary condition) and a dynamic force sensor with an applied static precompression and supported by a low-friction Teflon surface mount. A frequency sweep is performed by applying a sinusoidal displacement excitation $U\sin(\omega t)$  ($U$ is amplitude and $\omega$ is frequency) on the left side (undamped resonator/mass). The exerted force $F_{out} \sin(\omega t+\alpha)$, ($F_{out}$ is force amplitude and $\alpha$ is phase lag) is measured on the right side for various coupling spring stiffnesses ($\kappa$). (c) The eigenmodes of the non-Hermitian system are skewed and become degenerate at the exceptional point corresponding to a critical coupling $\kappa=\kappa_{EP}$. (d) Theoretically predicted reaction force amplitude as a function of excitation frequency $(\omega)$ and coupling $(\kappa)$. The force exerted on the force sensor is twice as large in the proximity of the EPD (red dot) as compared to when metamaterial is operating away from the EPD, i.e., with higher coupling spring stiffness $(\kappa>>\kappa_{EP})$.}
	\label{her1}
\end{figure}

\begin{figure}[t]
	\centering
	\includegraphics[width=0.9\textwidth]{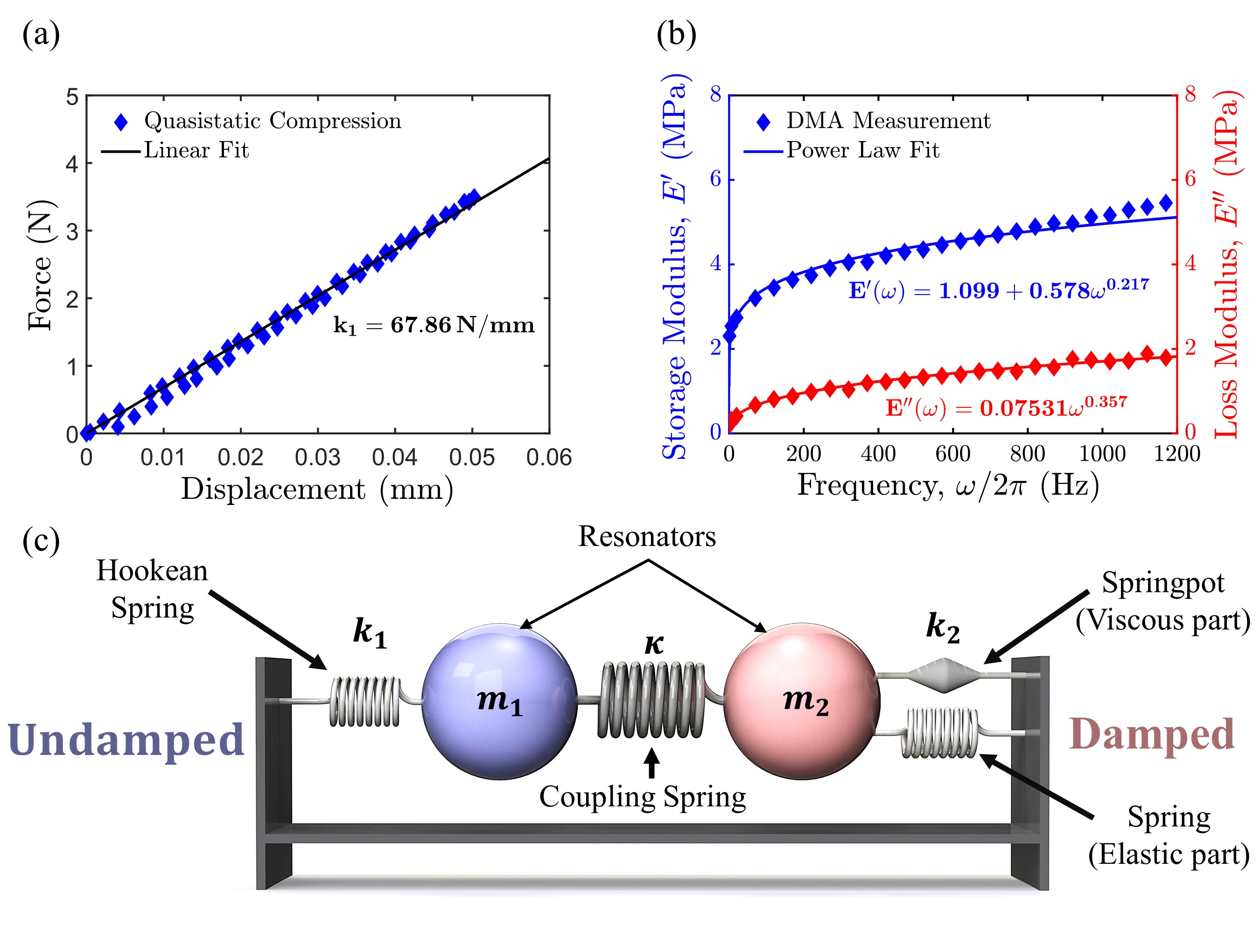}	\caption{(a) Force vs. Displacement of the compliant aluminum Hookean spring $(F=k_1 x_1)$. (b) Dynamic mechanical analysis (DMA) of frequency-dependent storage and loss moduli of PDMS and corresponding power law fits. (c) Conceptual design of non-Hermitian coupled resonators with differential damping under fixed boundary condition. Viscoelastic material (damped component) is represented by a parallel combination of an elastic spring and a springpot, where $k_2$ is the complex dynamic stiffness of the viscoelastic material. Undamped component is represented by a Hookean spring of stiffness $k_1$. Damped and undamped components are coupled together by a coiled spring of spring constant $\kappa$.}
	\label{her2}
\end{figure}

To this end, we have designed a two-mode cavity made by a non-Hermitian elastodynamic metamaterial consisting of two coupled resonant elements with differential damping (or non-proportional damping) between them (\Cref{her1}(a,b)). The damped $(m_2=m=4.635 \;g)$ and the undamped $(m_1=m)$ resonators (masses) are coupled by a coiled Hookean spring of stiffness $\kappa$, see \Cref{her1}(a,b). When the Hookean coupling strength between the two resonators acquires a critical value, the system exhibits an EPD associated with a coalescence of its eigenmodes (\Cref{her1}(c)) and the corresponding eigenfrequencies (\Cref{her1}(d)). Beyond this critical coupling, the eigenfrequencies bifurcate into two resonant modes with approximately the same constant linewidths (\Cref{her1}(d)). Linewidth is a measure of the spectral width of a resonance peak---characterized by the half-width of the resonance force peak at $1/ \sqrt{2}$ of the peak's height. It is equal to the imaginary part of the complex eigenfrequency. In this domain, a constant-amplitude dynamic force applied on the undamped resonator side of the metamaterial results in an actuation force—i.e., the force exerted by the damped resonator side of the metamaterial onto the dynamic force sensor—whose amplitude is controlled by the coupling strength between the two masses without any deterioration of force-signal quality. The input force is enhanced by $6.4$ times at the proximity of an EPD compared to the $3.2$ times enhancement produced by the same cavity-source system when the metamaterial does not support an EPD, see \Cref{her1}(d) and Figure S2 in \hyperref[section:sd]{SI}. A relation between the actuation force and the dissipative power from the viscous element allows to employ a mapping between electrodynamics and elastodynamics: a monochromatic source (actuator) emits light into a \textquotedblleft dark\textquotedblright mode, i.e., non-radiative mode (undamped resonator), which is coupled to a \textquotedblleft bright\textquotedblright mode (damped resonator). At some critical coupling the emission power (dissipative power) from the \textquotedblleft bright\textquotedblright mode  experiences a four-fold enhancement, related to a two-fold enhancement of the exerted actuation force. This phenomenon cannot be explained by conventional Purcell physics as the cavity quality factor remains constant. Its origin is traced back to the coalescence of super-modes (eigenvectors of the coupled resonators system) near the EPD. It is reflected in a narrowing of the Green’s function that describes the metamaterial---and consequently its LDoS---which acquires a square Lorentzian line-shape whose peak has a fourfold enhancement with respect to the peak of a single resonance. Our proof-of-principle metamaterial system promotes a new class of nano-indenters, atomic force microscopes and robotic-actuators.

To achieve differential damping between the two resonant elements of the metamaterial, we have designed and fabricated two types of springs: a compliant aluminum spring with specifically designed stiffness for the undamped component and a viscoelastic Polydimethylsiloxane (PDMS) for the damped component (\Cref{her2}(a,b)). For the undamped component, we fabricated a lightweight aluminum spring (T6-6061), whose geometry was designed using finite element modeling such that the spring does not yield plastically under static pre-stress. Its effective stiffness allows for an EPD to form as we vary the coupling between the two elements. Quasistatic and dynamic mechanical analysis of the aluminum spring (\Cref{her2}(a)) indicates that it follows Hooke’s law with a spring constant $k_1=67.86\; N/mm$ as designed, resulting in a natural frequency of $\omega_0=\sqrt{k_1/m}\approx 609 \;Hz$ for the undamped resonator. We fabricated the PDMS films with different amounts of crosslinking, which was controlled by varying the percentage of the curing-agent (5 to 10\% wt.) to tailor its dynamic modulus and loss-tangent. Then, $6\; mm$ diameter cylindrical specimens were cutout and preconditioned through five quasistatic compressive loading-unloading cycles up to $30\%$ strain to remove any potential variations in the constitutive response of the pristine material. We used an optimal cylindrical specimen—a softer PDMS with large loss-tangent identified from the material’s storage and loss moduli measured by dynamic mechanical analysis (DMA)—as the damped spring component of the metamaterial. The DMA was performed by first applying a $10\%$ static precompression strain to  the cylindrical PDMS specimens and then harmonically exciting them with a piezoelectric actuator (Physik Instrumente P841.10) at prescribed strain amplitudes and at frequencies ranging from $1\;Hz$ to $1200\; Hz$ while measuring the response force by a dynamic force sensor (PCB 208C01). The measured amplitude and phase lag between the actuation and the dynamic force signal were then used to calculate the frequency-dependent storage $(E' (\omega))$ and loss $(E'' (\omega))$ moduli. We precisely calibrated our custom-built DMA setup by measuring the damping properties of different materials with known loss-tangents such that any phase lag occurring due to the force sensor and the actuator at higher frequencies is corrected for in the measurements. Also, the actuator and the force sensor in our setup are orders of magnitude stiffer than the materials and metamaterial we tested. Hence, any effect due to their deformation on the measurements is negligible. The measured storage and loss moduli were then fitted using power-law functions (see \Cref{her2}(b)). The fits were used for the numerical modeling as well as the design of the metamaterial. 
We model the dynamics of the metamaterial by the following integro-differential equations

\begin{equation}\label{eq1}
 m_{1} \ddot{x}_{1}(t)=-k_{1}\left[x_{1}(t)-x_{0}(t)\right]-\kappa\left[x_{1}(t)-x_{2}(t)\right] 
\end{equation}
\begin{equation}\label{eq2}
m_{2} \ddot{x}_{2}(t)=-\sigma(t) A-\kappa\left[x_{2}(t)-x_{1}(t)\right] 
\end{equation}
Assuming $E(t)$ is the relaxation modulus and $\epsilon(t)$ is the axial strain in PDMS, then $\sigma(t)$ is given by the Boltzmann superposition integral as follows

\begin{equation}
\sigma(t)=\int_{0}\limits^{t}  E(t-\tau) \frac{d \epsilon(\tau)}{d \tau} d\tau  
\end{equation}

Above equations describes two coupled masses $m_1$, $m_2$  with displacements $x_1 (t)$, $x_2 (t)$ respectively. The Hookean spring constant of the aluminum spring is $k_1$, and $x_0(t)=U\sin(\omega t)$ is the displacement prescribed by the actuator, while  $F_{\rm out}=-\sigma (t)A$ is the force exerted on the mass $m_2$ with $A(=28.274\;mm^2)$ and $\sigma(t)$ being the cross-sectional area and the stress response of the PDMS specimen respectively. After taking the Fourier transform, Eqs. (\ref{eq1},\ref{eq2}) in the frequency domain takes the form
\begin{equation}\label{eq3}
-m \omega^{2} x_{1}(\omega)+k_{1} x_{1}(\omega)+\kappa\left[x_{1}(\omega)-x_{2}(\omega)\right]=k_{1} U 
\end{equation}
\begin{equation}\label{eq4}
-m \omega^{2} x_{2}(\omega)+\kappa\left[x_{2}(\omega)-x_{1}(\omega)\right]+\sigma (\omega)A=0
\end{equation}

where $\sigma(\omega)$ is the dynamic stress amplitude which can be expressed as a function of frequency dependent storage $(E')$ and loss modulus $(E'')$ as follows

\begin{equation}
\sigma(\omega)= [E'(\omega)+iE''(\omega)]\epsilon(\omega) \;\;\;\;,\; i=\sqrt{-1}
\end{equation}

where $\epsilon(\omega)=x_2(\omega)/h$ is the dynamic strain amplitude and $h(=1.948\;mm)$ is the height of the PDMS specimen. Assuming $k_2(\omega)=(A/h)[E'(\omega)+iE''(\omega)]$ is the effective dynamic stiffness, Eq. (\ref{eq4}) can be rewritten as follows

\begin{equation}\label{eq4a}
-m \omega^{2} x_{2}(\omega)+\kappa\left[x_{2}(\omega)-x_{1}(\omega)\right]+ k_2(\omega)x_2(\omega)=0
\end{equation}

In our metamaterial design, the real part of the effective stiffness of the PDMS specimen $(E'(\omega)(A/h))$ is approximately equal to the Hookean spring's stiffness $(k_1)$ in the frequency range of our experiment, which makes the natural frequency of both damped and undamped oscillators (when they are not coupled together) roughly the same (zero detuning). The imaginary part $(E''(\omega)(A/h))$ behaves similarly to the real part because they are causally linked via the Kramers-Kronig relations due to their common origin---Fourier transform of a time-dependent relaxation modulus \cite{lakes2009viscoelastic}. Both storage and loss modulus stay almost constant in the frequency range of our experiment (\Cref{her2}(b)), thus keeping the differential damping constant. In systems with an independent source of dissipation, the damping factor can be tuned to tailor the location of the exceptional point \cite{fang2021universal,gupta2022requisites}. However, in our case, damping $(E'')$ and stiffness $(E')$ are connected. Thus, our system exhibits constant differential damping and zero detuning with coupling as the tunable parameter.

\begin{figure}[t]
	\centering
	\includegraphics[width=0.9\textwidth]{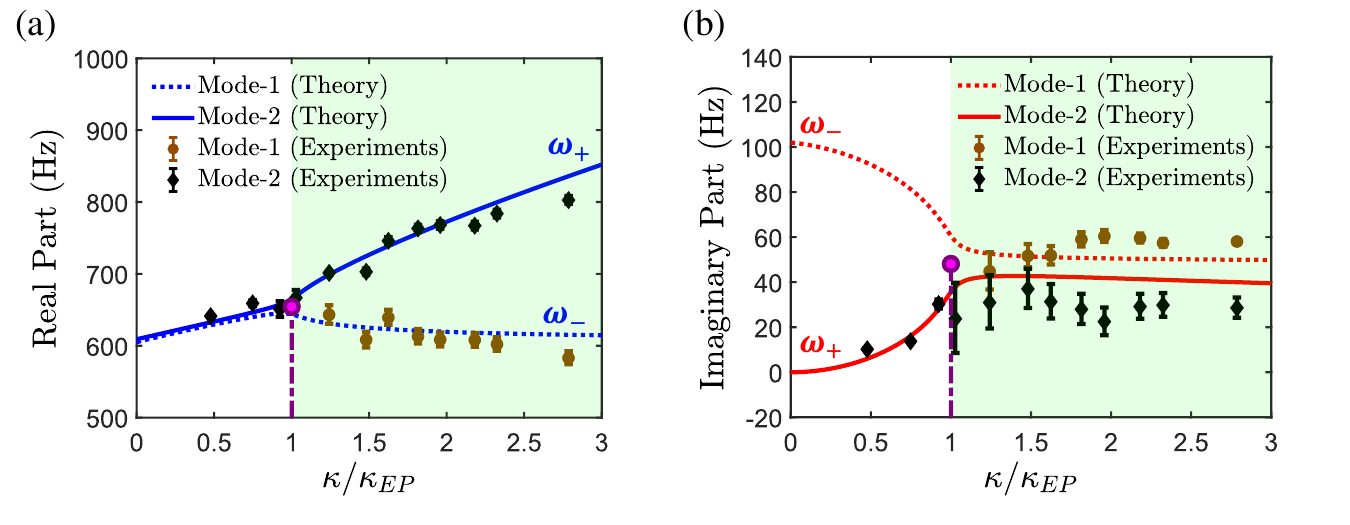}	\caption{ (a) The real part (peak position) and (b) the imaginary part (half of peak width at $1/\sqrt{2}$ of peak height) of resonant frequencies $(\omega_\pm)$ vs. coupling strength, obtained by a modal curve-fitting of the experimentally measured force spectra (\Cref{her4}(a)). The solid and dotted lines are the results of the simulations with the model described by Eqs. (\ref{eq3},\ref{eq4}) where dispersion effects are considered. The Hookean spring constant was taken to be $k_1=67.86\;N/mm$.}
	\label{her3}
\end{figure}

Eqs. (\ref{eq3},\ref{eq4a}) can be written in matrix form as follows

\begin{equation}
\left\{-\omega^2 \underbrace{\left[\begin{array}{cc}
m & 0 \\
0 & m
\end{array}\right]}_{\mathbf{M}}+\underbrace{\left[\begin{array}{cc}
k_1+{\kappa} & -\kappa \\
-\kappa & {\kappa}+\mathcal{R} e\left({k}_2(\omega)\right)
\end{array}\right]}_{\mathbf{K}}+i \underbrace{\left[\begin{array}{cc}
0 & 0 \\
0 & \mathcal{I}m\left({k}_2(\omega)\right)
\end{array}\right]}_{\mathbf{C}}\right\}\left\{\begin{array}{l}
{x}_1(\omega) \\
{x}_2(\omega)
\end{array}\right\}=\left\{\begin{array}{c}
{k}_1 U \\
0
\end{array}\right\} 
\end{equation}

where $\mathbf{M}$, $\mathbf{K}$, and $\mathbf{C}$ are the mass, stiffness, and damping matrices. Noteworthily, the damping matrix $(\mathbf{C})$ doesn't commute with the stiffness $(\mathbf{K})$ and mass $(\mathbf{M})$ matrices, which suggests that our system exhibits non-proportional damping \cite{phani2003necessary}. The corresponding eigenfrequencies $\omega_{\pm}$ of the metamaterial are found by solving the characteristic equation associated with Eqs. (\ref{eq3},\ref{eq4a}). \Cref{her3}(a,b) shows the real and imaginary parts of $\omega_{\pm}$ as functions of the stiffness $\kappa$ of the coiled coupling spring. A critical coupling, $\kappa_{EP}$ enforces an EPD, a coalescence of the resonant frequencies, $\omega_{-} = \omega_{+} = \omega_{EP}$, as well as the corresponding eigenvectors. As the coupling stiffness $\kappa$  is increased above $\kappa_{EP}$ the modes separate from one another, i.e. $\omega_{-}^{R} \equiv \mathcal{R}e\left(\omega_{-}\right) \neq \mathcal{R}e\left(\omega_{+}\right) \equiv \omega_{+}^{R}$, while their linewidths remain approximately the same, i.e. $\omega_-^I \equiv \mathcal{I}m(\omega_-)\approx \mathcal{I}m(\omega_+)\equiv\omega_+^I$ (see methods in \hyperref[section:sd]{SI}). Our modeling is further validated by a direct comparison with the measured eigenvalues (see methods in \hyperref[section:sd]{SI}), see \Cref{her3}(a,b). 

We further evaluate the mechanical power dissipated from mass $m_2$, which takes the form (see methods in \hyperref[section:sd]{SI})

\begin{equation}\label{eq5}
P_{d}(\omega)=\frac{\omega}{2 \pi} \mathrm{W}_{\mathrm{d}}=\Phi _{e} \cdot \xi (\omega) \;\;\;\;\;  , \; \xi(\omega)=\frac{2 \omega}{\pi} \mathcal{I}m( G_{11}(\omega))
\end{equation}

where $W_{d}(\omega)=A h \int_{0}^{2 \pi / \omega}  \sigma(t) \frac{d \epsilon}{d t} dt=(\pi A h) E^{\prime \prime}(\omega)|\epsilon(\omega)|^{2}$ is the dissipated energy per cycle \cite{lakes2009viscoelastic}, $\Phi_{e}=\frac{\pi}{4} \frac{F_{i n}^{2}}{m}$ depends only on the equivalent input force from the actuator $F_{in}=k_1 U$ acting on the mass $m_1=m$, and $G_{11} (\omega)$ is the element of the Green’s function that describes the system of Eqs. (\ref{eq3},\ref{eq4})

The Eq. (\ref{eq5}) establishes a connection between the power dissipated by the elastodynamic metamaterial shown in \Cref{her1}(a,b) and the Purcell physics of quantum optics. The analogies become clearer once we interpret the imaginary part of the Green’s function $\xi(\omega)$ as the LDoS of the mechanical resonant cavity formed by the metamaterial, and the piezoelectric actuator that excites the system as an emitter placed in the proximity of this cavity. Then the mechanical power emitted by the metamaterial is the analogue of the emissive power in the quantum optics framework.  In the latter framework, Purcell realized that the spontaneous emission rate of a quantum emitter is enhanced (or suppressed) by appropriately engineering its surrounding environment, and therefore, the LDoS \cite{purcell1946resonance}. 

Further theoretical progress can be made by assuming that the storage and loss moduli take a constant value in the frequency range of our experiments which are $E'=\frac{k_1 h}{A}=4.675\;MPa$ and $E''= 1.5235 \;MPa$ (see methods in \hyperref[section:sd]{SI}). This approximation is justified by the slow frequency dependence that the PDMS exhibits, see \Cref{her2}(b). Under this approximation, the critical coupling is $\kappa_{EP}=\frac{A}{2h} E''$, and for $\kappa \geq \kappa_{EP}$ the two resonant modes exhibit approximately the same decay rate $2 \omega_{\pm}^{I} \approx \Gamma \equiv \frac{A}{2 m h \omega_{0}} E^{\prime \prime}$ (\Cref{her3}(b)). In this case $\xi(\omega)$ becomes

\begin{equation}\label{eq6}
\xi(\omega)=\frac{1}{\pi} \frac{\Gamma \cdot \omega_{0} \cdot(\kappa / m)^{2} \omega}{\left(\omega^{2}-\omega_{0}^{2}\right)^{2}\left(\omega^{2}-\omega_{0}^{2}-2 \kappa / m\right)^{2}+\left(2 \Gamma \omega_{0}\right)^{2}\left(\omega^{2}-\omega_{0}^{2}-\kappa / m\right)^{2}}
\end{equation}

where $\omega_0=\sqrt{k_1/m}$. Because at the proximity of the EPD the two eigenfrequencies degenerate at $\omega_{\pm}^R=\omega_{EP} \approx\omega_0 \sqrt{1+\Gamma/\omega_0} \approx \omega_0$   (see methods in \hyperref[section:sd]{SI}), one naturally expects that $\xi(\omega)$, and subsequently $P_d (\omega)$, will also be modified. To better understand the consequence of EPD on them, we first evaluate $\xi(\omega)$ in the vicinity of $\omega\approx\omega_{EP}$. From Eq. (\ref{eq6}), we get that (see \hyperref[section:sd]{SI})

\begin{equation}\label{eq7}
\xi_{\mathrm{EP}}(\omega) \approx \frac{1}{2 \pi} \frac{(\Gamma / 2)^{3}}{\left[\left(\omega-\omega_{\mathrm{EP}}\right)^{2}+(\Gamma / 2)^{2}\right]^{2}}
\end{equation}

corresponding to a square-Lorentzian—as opposed to the more traditional Lorentzian line-shape characterizing the LDoS away from an EPD. In the other limiting case of $\kappa>>\kappa_{EP}$, the two resonant frequencies $\omega_-^R\approx\omega_0$ and $\omega_+^R\approx\sqrt{\omega_0^2+(2\kappa/m)}$ (see methods in \hyperref[section:sd]{SI}) are well separated from each other. In this case one can approximate Eq. (\ref{eq6}) as a sum of two Lorentzians centered at $\omega_{\pm}$ and having the same linewidth $\Gamma$. Specifically, we have

\begin{equation}\label{eq8}
\xi_{\infty}(\omega) \approx \frac{1}{8 \pi} \sum_{\pm} \frac{\Gamma / 2}{\left(\omega-\omega_{\pm}^{R}\right)^{2}+(\Gamma / 2)^{2}}    
\end{equation}

which can be used as a reference for measuring the effects of EPD in the power emission. At this point, it is instructive to introduce the rescaled dissipated power $\mathcal{P}\left(\omega_{e}\right)=P_{d}\left(\omega_{e}\right) / P_{d}^{\infty}\left(\omega_{\pm}^{R}\right)$ where $P_{d}^{\infty}\left(\omega_{\pm}\right) \equiv \frac{\pi}{2} \Phi_{e} \cdot \xi_{\infty}\left(\omega_{\pm}\right)$ and $\omega_{e}=\omega_{\pm}^{R}$ is the frequency of the actuation force which is considered to be monochromatic and at resonant frequency. Substituting Eqs. (\ref{eq7}) and (\ref{eq8}) back in the general expression for the emissivity Eq. (\ref{eq5}), we deduce that $\mathcal{P}(\omega_e=\omega_{EP} )=4$. A more careful analysis (see \hyperref[section:sd]{SI}) indicates that $\mathcal{P}(\omega_e=\omega_{EP})$ might be slightly larger/smaller than a factor of four, i.e., $\mathcal{P}(\omega_e=\omega_{EP})\approx 4+\mathcal{O}\left(\frac{\Gamma}{\omega_{0}}\right)+\mathcal{O}\left(\frac{\kappa}{k_{1}}\right)$, where $\frac{\Gamma}{\omega_0} << 1$; $\frac{\kappa}{k_1} <1$.

\begin{figure}[t]
	\centering
	\includegraphics[width=0.9\textwidth]{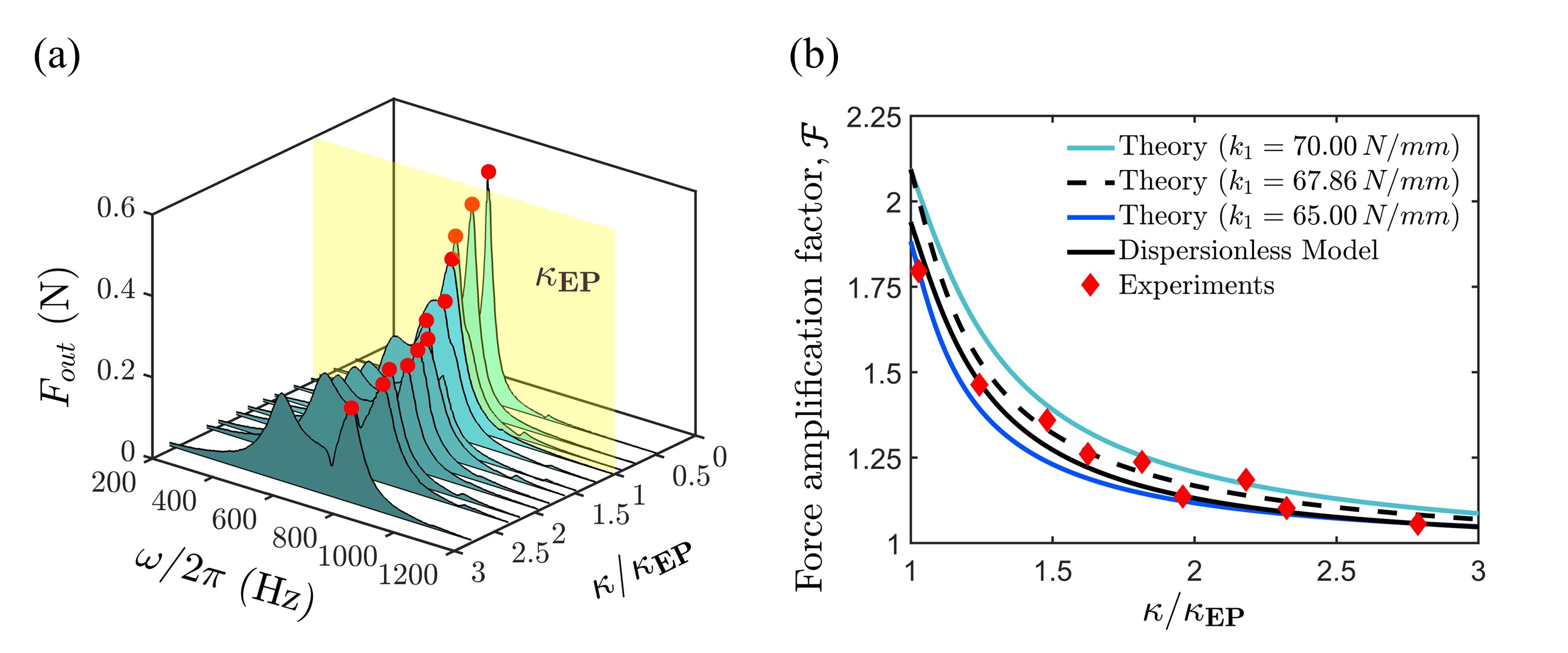}	\caption{(a) Frequency-swept force amplitude measured experimentally for various coupling spring stiffnesses. Splitting of a single sharp peak into two broad peaks can be observed as coupling increases. (b) Emitted force amplification factor $\mathcal{F}(\omega_\pm)$ vs. the rescaled coupling. The experimental data are indicated in filled diamonds. The solid black line indicates the results of the dispersionless modeling with $k_1=70\;N/mm$. Black dashed line indicate the results of the simulation of model Eqs. (\ref{eq3},\ref{eq4}) for $k_1=67.86\;N/mm$---corresponding to best fit model. Curves corresponding to $k_1=70\;N/mm$ and $k_1=65\;N/mm$ represent the upper and lower uncertainty bounds of data fitting in \Cref{her2}. A two-fold enhancement is observed in all cases.}
	\label{her4}
\end{figure}

The enhanced dissipated power due to the four-fold enhancement of the peak of $\xi(\omega)$ is also reflected in the actuation force emitted from the metamaterial and measured by the dynamic force sensor attached to the $m_2$-resonator. The latter is written in the form $F_{\rm out} (t)=F_{\rm out} (\omega_e)\sin(\omega_{e} t+\alpha$), where $\omega_e$ is the driving frequency of the actuator. At the same time, the emitted force amplitude  $|F_{\rm out} (\omega_e)|$ is 

\begin{equation}\label{eq9}
\left|F_{\text {out }}\left(\omega_{e}\right)\right| \equiv \mathrm{A}\left|\sigma\left(\omega_{e}\right)\right|=\sqrt{\frac{2 A\left(\left|E^{\prime}\left(\omega_{e}\right)\right|^{2}+\left|E^{\prime \prime}\left(\omega_{e}\right)\right|^{2}\right)}{h \omega_{e} E^{\prime \prime}\left(\omega_{e}\right)} P_{d}\left(\omega_{e}\right)}
\end{equation}

where we have expressed the stress function as $\sigma(\omega)=[E'(\omega)+iE''(\omega)]\epsilon(\omega)$ and subsequently we have used Eq. (\ref{eq5}) to express the strain in terms of the emitted power as $|\epsilon(\omega)|=\sqrt{4 P_{d}(\omega) /\left(\pi E^{\prime \prime} A h\right)}$. From Eq. (\ref{eq9}) we can further estimate the force amplification factor $\mathcal{F}(\omega_e)\equiv\frac{F_{\text {out }}\left(\omega_{e}\right)}{F_{\text {out }}^{\infty}\left(\omega_{\pm}^{R}\right)}=\sqrt{\mathcal{P}\left(\omega_{e}\right)}$, which for $\kappa_{EP}$  and driving frequency $\omega_e=\omega_{EP}$ is $\mathcal{F}(\omega_e=\omega_{EP} )=2$.  We point out that the force amplification factor $\mathcal{F}(\omega_e )$ indicates an additional enhancement relative to the typical Purcell force enhancement occurring in the actuating force $F_{\rm out}^{\infty} (\omega_{\pm}^R )$ (see Figure S2 in \hyperref[section:sd]{SI}) for the same metamaterial-source configuration when the metamaterial does not support an EPD.

We have tested the validity of the above predictions by direct measurements of the emitted actuation force amplitude (\Cref{her4}(a)) for various coupling stiffness $\kappa$ versus the driving frequency $\omega_e$ (see methods in \hyperref[section:sd]{SI}). From these data we extracted the force amplification factor $\mathcal{F}(\omega_e=\omega_\pm)$ and report it as a function of the coupling stiffness $\kappa$ normalized by the coupling stiffness corresponding to the EPD, $\kappa_{EP}$, in \Cref{her4}(b). The experimental data confirm the above theoretical predictions very well. In the same figure, we also report the force amplification factor calculated numerically from Eqs. (\ref{eq3},\ref{eq4}). From these equations, we extracted the frequency-dependent displacement amplitude $x_2 (\omega)$ using the dispersion characteristics of the stress function, which is related to the strain in viscoelastic material as $\epsilon(\omega)=x_2 (\omega)/h$. Using the definition of $|F_{\rm out} (\omega)|$ in Eq. (\ref{eq9}), we have calculated $\mathcal{F}(\omega_e=\omega_{\pm}^R )$ for three different $k_1$ constants to evaluate the robustness of the observed two-fold enhancement. In the same figure, we also report the theoretical results of the non-dispersive model (solid black line). All curves confirm the two-fold enhancement in exerted actuation force at the proximity of the EPD with respect to the actuation force value corresponding to large coupling constants where a standard cavity-enhancement Purcell effect occurs.

In conclusion, we have demonstrated experimentally and analyzed theoretically an enhancement in the exerted actuation force and the emitted mechanical power by a non-Hermitian elastodynamic cavity---a metamaterial consisting of undamped and damped resonators. It occurs when the actuation source is coupled only to the undamped resonator (a \textquotedblleft dark\textquotedblright mode) while the emission is measured on the viscoelastic element (the \textquotedblleft bright\textquotedblright mode). The origin of this enhancement has been traced to the reorganization of the super-modes of the metamaterial in the proximity of an EPD. Our experiments pave the way for understanding the ramifications of non-Hermiticity and EPD engineering in the manipulation of the Purcell factor of narrow-band emitters. Additionally, while coupled oscillators and periodic structures have been extensively studied in vibroacoustics under various scenarios, our work provides a unique metamaterial design pathway that utilizes differential damping as a design element to realize exceptional point degeneracies experimentally and demonstrates a novel phenomenon of enhanced emission at constant quality factor. Our work paves the way to an unexplored realm of physics in non-Hermitian systems and their potential in various applications, such as a new class of indenters for material hardness measurements \cite{broitman2017indentation} or enhanced reconfigurable actuators in robotics.

\section*{Acknowledgements}

RT and AG acknowledge the financial support from the Solid Mechanics Program of the Army Research Office (ARO) (Award No.: W911NF2010160) and the Dynamics, Control, and System Diagnostics (DCSD) Program of the National Science Foundation (NSF) (Award No.: NSF-CMMI-1925530). TK and AK acknowledge the financial support from the DCSD Program of the NSF (Award No.: NSF-CMMI-1925543) and from the Simons Foundation for Collaboration in MPS grant No 733698. We also acknowledge the assistance of Dr. Jizhe Cai on PDMS sample fabrication

\section*{Appendix A. Supplementary data}

Supplementary material related to this article can be found
online

\label{section:sd}
\bibliography{mybibfile}

\begin{thebibliography}{10}
\expandafter\ifx\csname url\endcsname\relax
  \def\url#1{\texttt{#1}}\fi
\expandafter\ifx\csname urlprefix\endcsname\relax\def\urlprefix{URL }\fi
\expandafter\ifx\csname href\endcsname\relax
  \def\href#1#2{#2} \def\path#1{#1}\fi

\bibitem{liu2000locally}
Z.~Liu, X.~Zhang, Y.~Mao, Y.~Zhu, Z.~Yang, C.~T. Chan, P.~Sheng, Locally
  resonant sonic materials, science 289~(5485) (2000) 1734--1736.

\bibitem{hussein2014dynamics}
M.~I. Hussein, M.~J. Leamy, M.~Ruzzene, Dynamics of phononic materials and
  structures: Historical origins, recent progress, and future outlook, Applied
  Mechanics Reviews 66~(4).

\bibitem{boechler2011bifurcation}
N.~Boechler, G.~Theocharis, C.~Daraio, Bifurcation-based acoustic switching and
  rectification, Nature materials 10~(9) (2011) 665--668.

\bibitem{zhang2011broadband}
S.~Zhang, C.~Xia, N.~Fang, Broadband acoustic cloak for ultrasound waves,
  Physical review letters 106~(2) (2011) 024301.

\bibitem{cha2018experimental}
J.~Cha, K.~W. Kim, C.~Daraio, Experimental realization of on-chip topological
  nanoelectromechanical metamaterials, Nature 564~(7735) (2018) 229--233.

\bibitem{nassar2017non}
H.~Nassar, H.~Chen, A.~Norris, M.~Haberman, G.~Huang, Non-reciprocal wave
  propagation in modulated elastic metamaterials, Proceedings of the Royal
  Society A: Mathematical, Physical and Engineering Sciences 473~(2202) (2017)
  20170188.

\bibitem{bender2007making}
C.~M. Bender, Making sense of non-hermitian hamiltonians, Reports on Progress
  in Physics 70~(6) (2007) 947.

\bibitem{kato2013perturbation}
T.~Kato, Perturbation theory for linear operators, Vol. 132, Springer Science
  \& Business Media, 2013.

\bibitem{el2018non}
R.~El-Ganainy, K.~G. Makris, M.~Khajavikhan, Z.~H. Musslimani, S.~Rotter, D.~N.
  Christodoulides, Non-hermitian physics and pt symmetry, Nature Physics 14~(1)
  (2018) 11--19.

\bibitem{miri2019exceptional}
M.-A. Miri, A.~Al{\`u}, Exceptional points in optics and photonics, Science
  363~(6422) (2019) eaar7709.

\bibitem{thomas2016giant}
R.~Thomas, H.~Li, F.~M. Ellis, T.~Kottos, Giant nonreciprocity near
  exceptional-point degeneracies, Physical Review A 94~(4) (2016) 043829.

\bibitem{doppler2016dynamically}
J.~Doppler, A.~A. Mailybaev, J.~B{\"o}hm, U.~Kuhl, A.~Girschik, F.~Libisch,
  T.~J. Milburn, P.~Rabl, N.~Moiseyev, S.~Rotter, Dynamically encircling an
  exceptional point for asymmetric mode switching, Nature 537~(7618) (2016)
  76--79.

\bibitem{assawaworrarit2017robust}
S.~Assawaworrarit, X.~Yu, S.~Fan, Robust wireless power transfer using a
  nonlinear parity--time-symmetric circuit, Nature 546~(7658) (2017) 387--390.

\bibitem{xu2016topological}
H.~Xu, D.~Mason, L.~Jiang, J.~Harris, Topological energy transfer in an
  optomechanical system with exceptional points, Nature 537~(7618) (2016)
  80--83.

\bibitem{shi2016accessing}
C.~Shi, M.~Dubois, Y.~Chen, L.~Cheng, H.~Ramezani, Y.~Wang, X.~Zhang, Accessing
  the exceptional points of parity-time symmetric acoustics, Nature
  communications 7~(1) (2016) 1--5.

\bibitem{thevamaran2019asymmetric}
R.~Thevamaran, R.~M. Branscomb, E.~Makri, P.~Anzel, D.~Christodoulides,
  T.~Kottos, E.~L. Thomas, Asymmetric acoustic energy transport in
  non-hermitian metamaterials, The Journal of the Acoustical Society of America
  146~(1) (2019) 863--872.

\bibitem{peng2016anti}
P.~Peng, W.~Cao, C.~Shen, W.~Qu, J.~Wen, L.~Jiang, Y.~Xiao, Anti-parity--time
  symmetry with flying atoms, Nature Physics 12~(12) (2016) 1139--1145.

\bibitem{lee2015macroscopic}
J.~Lee, T.~Kottos, B.~Shapiro, Macroscopic magnetic structures with balanced
  gain and loss, Physical Review B 91~(9) (2015) 094416.

\bibitem{zhang2017observation}
D.~Zhang, X.-Q. Luo, Y.-P. Wang, T.-F. Li, J.~You, Observation of the
  exceptional point in cavity magnon-polaritons, Nature communications 8~(1)
  (2017) 1--6.

\bibitem{dominguez2020environmentally}
V.~Dom{\'\i}nguez-Rocha, R.~Thevamaran, F.~Ellis, T.~Kottos, Environmentally
  induced exceptional points in elastodynamics, Physical Review Applied 13~(1)
  (2020) 014060.

\bibitem{fang2021universal}
Y.~Fang, T.~Kottos, R.~Thevamaran, Universal route for the emergence of
  exceptional points in pt-symmetric metamaterials with unfolding spectral
  symmetries, New Journal of Physics 23~(6) (2021) 063079.

\bibitem{lustig2019anomalous}
B.~Lustig, G.~Elbaz, A.~Muhafra, G.~Shmuel, Anomalous energy transport in
  laminates with exceptional points, Journal of the Mechanics and Physics of
  Solids 133 (2019) 103719.

\bibitem{rosa2021exceptional}
M.~I. Rosa, M.~Mazzotti, M.~Ruzzene, Exceptional points and enhanced
  sensitivity in pt-symmetric continuous elastic media, Journal of the
  Mechanics and Physics of Solids 149 (2021) 104325.

\bibitem{li2022experimental}
X.~Li, Z.~Yu, H.~Iizuka, T.~Lee, Experimental demonstration of extremely
  asymmetric flexural wave absorption at the exceptional point, Extreme
  Mechanics Letters 52 (2022) 101649.

\bibitem{fang2022emergence}
Y.~Fang, T.~Kottos, R.~Thevamaran, Emergence of exceptional points in periodic
  metastructures with hidden parity-time symmetric defects, Journal of Applied
  Mechanics 89~(12) (2022) 121003.

\bibitem{guo2009observation}
A.~Guo, G.~Salamo, D.~Duchesne, R.~Morandotti, M.~Volatier-Ravat, V.~Aimez,
  G.~Siviloglou, D.~Christodoulides, Observation of p t-symmetry breaking in
  complex optical potentials, Physical review letters 103~(9) (2009) 093902.

\bibitem{lin2011unidirectional}
Z.~Lin, H.~Ramezani, T.~Eichelkraut, T.~Kottos, H.~Cao, D.~N. Christodoulides,
  Unidirectional invisibility induced by p t-symmetric periodic structures,
  Physical Review Letters 106~(21) (2011) 213901.

\bibitem{hodaei2014parity}
H.~Hodaei, M.-A. Miri, M.~Heinrich, D.~N. Christodoulides, M.~Khajavikhan,
  Parity-time--symmetric microring lasers, Science 346~(6212) (2014) 975--978.

\bibitem{feng2014single}
L.~Feng, Z.~J. Wong, R.-M. Ma, Y.~Wang, X.~Zhang, Single-mode laser by
  parity-time symmetry breaking, Science 346~(6212) (2014) 972--975.

\bibitem{hodaei2017enhanced}
H.~Hodaei, A.~U. Hassan, S.~Wittek, H.~Garcia-Gracia, R.~El-Ganainy, D.~N.
  Christodoulides, M.~Khajavikhan, Enhanced sensitivity at higher-order
  exceptional points, Nature 548~(7666) (2017) 187--191.

\bibitem{chen2017exceptional}
W.~Chen, {\c{S}}.~Kaya~{\"O}zdemir, G.~Zhao, J.~Wiersig, L.~Yang, Exceptional
  points enhance sensing in an optical microcavity, Nature 548~(7666) (2017)
  192--196.

\bibitem{lai2019observation}
Y.-H. Lai, Y.-K. Lu, M.-G. Suh, Z.~Yuan, K.~Vahala, Observation of the
  exceptional-point-enhanced sagnac effect, Nature 576~(7785) (2019) 65--69.

\bibitem{hokmabadi2019non}
M.~P. Hokmabadi, A.~Schumer, D.~N. Christodoulides, M.~Khajavikhan,
  Non-hermitian ring laser gyroscopes with enhanced sagnac sensitivity, Nature
  576~(7785) (2019) 70--74.

\bibitem{kononchuk2022enhanced}
R.~Kononchuk, J.~Cai, F.~Ellis, R.~Thevamaran, T.~Kottos, Enhanced
  signal-to-noise performance of ep-based electromechanical accelerometers,
  arXiv preprint arXiv:2201.13328.

\bibitem{lin2016enhanced}
Z.~Lin, A.~Pick, M.~Lon{\v{c}}ar, A.~W. Rodriguez, Enhanced spontaneous
  emission at third-order dirac exceptional points in inverse-designed photonic
  crystals, Physical review letters 117~(10) (2016) 107402.

\bibitem{pick2017general}
A.~Pick, B.~Zhen, O.~D. Miller, C.~W. Hsu, F.~Hernandez, A.~W. Rodriguez,
  M.~Solja{\v{c}}i{\'c}, S.~G. Johnson, General theory of spontaneous emission
  near exceptional points, Optics express 25~(11) (2017) 12325--12348.

\bibitem{pick2017enhanced}
A.~Pick, Z.~Lin, W.~Jin, A.~W. Rodriguez, Enhanced nonlinear frequency
  conversion and purcell enhancement at exceptional points, Physical Review B
  96~(22) (2017) 224303.

\bibitem{ren2021quasinormal}
J.~Ren, S.~Franke, S.~Hughes, Quasinormal modes, local density of states, and
  classical purcell factors for coupled loss-gain resonators, Physical Review X
  11~(4) (2021) 041020.

\bibitem{franke2021fermi}
S.~Franke, J.~Ren, M.~Richter, A.~Knorr, S.~Hughes, Fermi’s golden rule for
  spontaneous emission in absorptive and amplifying media, Physical Review
  Letters 127~(1) (2021) 013602.

\bibitem{zhong2021control}
Q.~Zhong, A.~Hashemi, {\c{S}}.~{\"O}zdemir, R.~El-Ganainy, Control of
  spontaneous emission dynamics in microcavities with chiral exceptional
  surfaces, Physical Review Research 3~(1) (2021) 013220.

\bibitem{khanbekyan2020decay}
M.~Khanbekyan, J.~Wiersig, Decay suppression of spontaneous emission of a
  single emitter in a high-q cavity at exceptional points, Physical Review
  Research 2~(2) (2020) 023375.

\bibitem{hashemi2022linear}
A.~Hashemi, K.~Busch, D.~Christodoulides, S.~Ozdemir, R.~El-Ganainy, Linear
  response theory of open systems with exceptional points, Nature
  Communications 13~(1) (2022) 1--12.

\bibitem{purcell1946resonance}
E.~M. Purcell, H.~C. Torrey, R.~V. Pound, Resonance absorption by nuclear
  magnetic moments in a solid, Physical review 69~(1-2) (1946) 37.

\bibitem{dirac1927quantum}
P.~A.~M. Dirac, The quantum theory of the emission and absorption of radiation,
  Proceedings of the Royal Society of London. Series A, Containing Papers of a
  Mathematical and Physical Character 114~(767) (1927) 243--265.

\bibitem{betzig1993single}
E.~Betzig, R.~J. Chichester, Single molecules observed by near-field scanning
  optical microscopy, Science 262~(5138) (1993) 1422--1425.

\bibitem{muhlschlegel2005resonant}
P.~Muhlschlegel, H.-J. Eisler, O.~J. Martin, B.~Hecht, D.~Pohl, Resonant
  optical antennas, science 308~(5728) (2005) 1607--1609.

\bibitem{khajavikhan2012thresholdless}
M.~Khajavikhan, A.~Simic, M.~Katz, J.~Lee, B.~Slutsky, A.~Mizrahi, V.~Lomakin,
  Y.~Fainman, Thresholdless nanoscale coaxial lasers, Nature 482~(7384) (2012)
  204--207.

\bibitem{choy2011enhanced}
J.~T. Choy, B.~J. Hausmann, T.~M. Babinec, I.~Bulu, M.~Khan, P.~Maletinsky,
  A.~Yacoby, M.~Lon{\v{c}}ar, Enhanced single-photon emission from a
  diamond--silver aperture, Nature Photonics 5~(12) (2011) 738--743.

\bibitem{schmidt2018elastic}
M.~K. Schmidt, L.~Helt, C.~G. Poulton, M.~Steel, Elastic purcell effect,
  Physical Review Letters 121~(6) (2018) 064301.

\bibitem{fink2013subwavelength}
M.~Fink, F.~Lemoult, J.~d. Rosny, A.~Tourin, G.~Lerosey, Subwavelength
  focussing in metamaterials using far field time reversal, in: Acoustic
  Metamaterials, Springer, 2013, pp. 141--168.

\bibitem{broitman2017indentation}
E.~Broitman, Indentation hardness measurements at macro-, micro-, and
  nanoscale: a critical overview, Tribology Letters 65~(1) (2017) 1--18.

\bibitem{lakes2009viscoelastic}
R.~S. Lakes, Viscoelastic materials, Cambridge university press, 2009.

\bibitem{gupta2022requisites}
A.~Gupta, R.~Thevamaran, Requisites on viscoelasticity for exceptional points
  in passive elastodynamic metamaterials, arXiv preprint arXiv:2209.04960.

\bibitem{phani2003necessary}
A.~S. Phani, On the necessary and sufficient conditions for the existence of
  classical normal modes in damped linear dynamic systems, Journal of Sound and
  Vibration 264~(3) (2003) 741--745.

\end{thebibliography}

\end{document}